\documentclass[twocolumn,twoside,slac_two]{revtex4}
\usepackage{graphicx}
\usepackage{fancyhdr}

\begin{document}

\title{Cherenkov and Jansky: Our Understanding of AGN at the Highest Energies}

\author{J. S. Perkins}
\affiliation{Center for Research and Exploration in Space Science and Technology}
\affiliation{Astroparticle Physics Laboratory NASA/GSFC, Greenbelt, MD 20771, USA}
\affiliation{University of Maryland, Baltimore County, 1000 Hilltop Circle, Baltimore, MD 21250, USA}
\author{on behalf of the VERITAS and {\em Fermi} LAT Collaborations}
\affiliation{http://veritas.sao.arizona.edu}
\affiliation{http://www-glast.stanford.edu}

\begin{abstract}
Misaligned blazars have been the subject of some of the most
successful radio and gamma-ray multiwavelength campaigns.  These
campaigns have included many of the major ground and space based
gamma-ray telescopes and span decades of energy.  Even though
misaligned blazars account for only a small number of the total AGN
detected at VHE, they provide a unique view on the AGN population.  By
viewing blazars at larger angles to our line of sight, they become a
unique laboratory for the study of AGN jet substructure and the
morphology of non-thermal emission processes.  This contribution will
discuss our understanding of three VHE misaligned blazars.
\end{abstract}

\maketitle

\thispagestyle{fancy}

\section{Introduction}

\begin{figure*}[t]
\centering
\includegraphics[width=110mm]{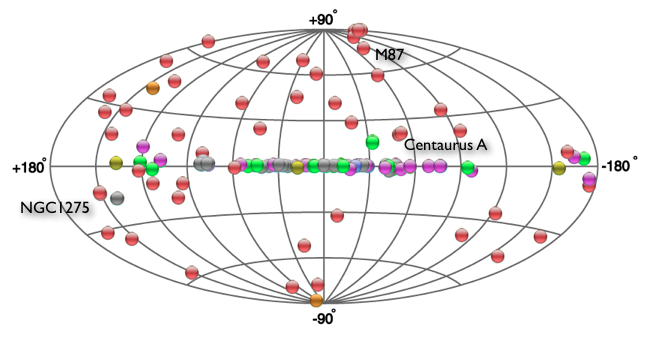}
\caption{As of this meeting (November 2011) There are currently over
  120 VHE sources, about 40 of which are blazars (the red dots shown
  in this figure).  This contribution focuses on the three misaligned
  AGN highlighted in the above figure.  Figure from {\em
    http://tevcat.uchicago.edu}.}
\label{fig:skymap}
\end{figure*}

$\gamma$-ray astronomy has made major advances in the last decade.
The construction and operation of the current generation of very high
energy (VHE; E $>$ 100 GeV) instruments like VERITAS, HESS and MAGIC
have resulted in the detection of more than 120 VHE sources (see
Figure \ref{fig:skymap}).  The launch of the {\it Fermi} satellite
\citep{Atwood2009} has opened new windows on the high energy (HE; 100
MeV $<$ E $<$ 100 GeV) sky.  Individually, these instruments have
produced many good results, but the fact that they can perform
multiwavelength studies increases their scientific output
dramatically.  It is important to remember that all of the work
presented here is the result of collaborations between many excellent
people, not only in VERITAS and {\em Fermi} but in MAGIC and HESS and
many, many radio scientists.

The field of VHE astronomy is necessarily ground based (see
\citet{Hinton2009a} for a review).  You cannot get enough effective
area at these energies with a space based instrument, while the ground
based techniques result in an effective area the size of a football
field for the current generation of telescopes. These large effective
areas are possible due to the Cherenkov technique employed by the VHE
community. There are three major arrays in use right now: VERITAS,
HESS and MAGIC and the field continues to move forward, there are
upgrades in progress (HESS2 \citep{Moudden2011}, VERITAS upgrade
\citep{KiedaD.B.2011}) and completed (MAGIC2 \citep{LombardiS.2011}).
Also, a new generation of instruments is coming online in the next
years (HAWC \citep{GonzalezM.M.2011}, CTA \citep{CTAConsortium2010}).

There are over 120 VHE sources in the catalog
(http://tevcat.uchicago.edu).  The most populous class of objects
(about 40) are blazars and a handful of these are misaligned blazars
(see Figure \ref{fig:doppler}).  It is important to note that all of
the misaligned blazars are also {\it Fermi} LAT sources, allowing a
simultaneous measurement of the $\gamma$-ray SED.  This is not the
place for a review of AGN physics but the basic understanding is that
these objects contain a central supermassive black hole surrounded by
an accretion disk that powers a relativistic jet of photons and
particles.  The orientation of this jet towards the observer
determines the source type: a jet pointed directly at the earth is
seen as a BL Lac object or FSRQ while one that is misaligned is seen
as some type of radio galaxy.  There are many open questions that we
are trying to answer by studying AGN at $\gamma$-ray wavelengths
including: determining the emission mechanisms, understanding the
accretion physics, and finding the emission location.

\section{Misaligned Blazars}

\begin{figure}
\includegraphics[width=65mm]{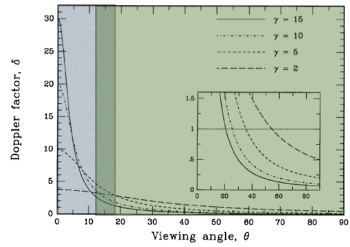}
\caption{Shows the Doppler factor versus the viewing angle for AGN for
  different Lorentz factors.  Radio galaxies are usually thought to
  occur in the green region while blazars are covered by the blue
  region (Figure from \citet{Urry1995}).}
\label{fig:doppler}
\end{figure}

Only a handful of the detected VHE AGN are misaligned.  The reason for
this can be readily understood by looking at Figure \ref{fig:doppler}.
Blazar emission benefits from high Doppler and Lorentz factors which
boost the flux and energy of the detected $\gamma$-rays and increases
the detection probability.  Jet emission is still possible at large
viewing angles but the Doppler factor will be small and thus these
types of objects are harder to detect at VHE.  It is thought that the
same emission mechanism that occurs in blazars is what is seen in
these misaligned AGN and so the same types of models could be used to
understand the observed SED \citep{Urry1995}.  Additionally, there is
also the possibility to see lobe emission from regions outside of the
core (as is seen in Centaurus A from the LAT \citep{Abdo2010h}).

The complete study of misaligned AGN is only possible through
multiwavelength studies.  The main reason for this is that these
sources emit photons throughout the electromagnetic spectrum and to
accurately understand the emission mechanisms, you need measurements
all the way from the radio to the $\gamma$-ray.  A complication is
that these sources are known to be highly variable and thus, you not
only need multiwavelength studies, but simultaneous multiwavelength
studies.  These are difficult, but the payoff is great since we can
learn about the AGN population as a whole by fitting these types of
objects into a general AGN emission model.  We learn new things about
the diverse AGN class by observing AGN of different types.

\subsection{NGC1275}

\begin{figure}
\includegraphics[width=65mm]{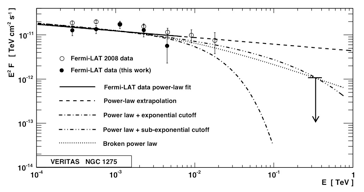}
\caption{VERITAS upper limit (99\% confidence) and LAT spectrum (solid
  points are for the time period of the VERITAS observations and open
  points are for the first year of LAT operations) of NGC 1275 with
  models.  The upper limit at VHE indicated that the spectrum is not
  well fit by a simple power law (Figure from \citet{Acciari2009a}).}
\label{fig:ngc1275spectrum}
\end{figure}

\begin{figure}
\includegraphics[width=65mm]{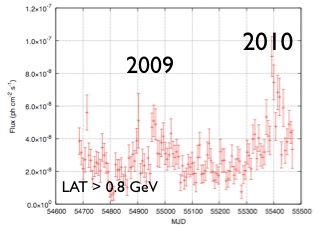}
\caption{LAT lightcurve of NGC1275.  The VHE detection by MAGIC
  occurred during the flare in 2010 (Figure from \citet{Brown2011}).}
\label{fig:ngc1275lightcurve}
\end{figure}

NGC 1275 is a radio galaxy at the core of the Perseus cluster.  It was
initially detected at GeV energies in three months of {\it Fermi}
observations \citep{Abdo2009a}.  This indicated strong evidence for
variability since it was detected by COS B \citep{StrongA.W.1982} but
not by EGRET.  Observations with VERITAS did not yield a detection at
VHE energies but the upper limit combined with the LAT measurements
indicated that the full SED is not compatible with a single power law
(\citet{Acciari2009a}, Figure \ref{fig:ngc1275spectrum}).

\begin{figure}
\includegraphics[width=65mm]{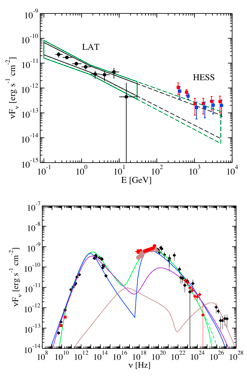}
\caption{The upper plot shows the spectra seen by the LAT (in black)
  and HESS (in blue and red).  The butterfly shows the overall fit of
  the LAT data extended to the HESS energy range.  In the lower plot,
  are model fits to the nuclear region of Cen A. The green curve is a
  synchrotron plus SSC fit to the entire data set. The dashed green
  curve shows this model without $\gamma$-$\gamma$ attenuation. The
  violet curve is a similar fit but is designed to under fit the X-ray
  data, and the brown curve is designed to fit the HESS data while not
  over-producing the other data in the SED. The blue curve is a
  decelerating jet model fit (Figures from \citet{Abdo2010c}).}
\label{fig:cenavhe}
\end{figure}

\begin{figure*}[t]
\centering
\includegraphics[width=135mm]{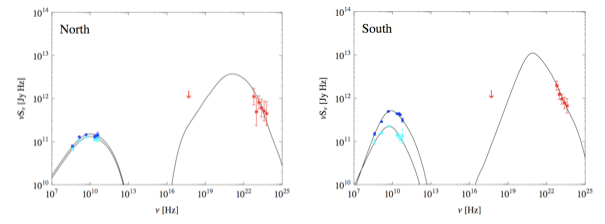}
\caption{The spectrum of the North and South lobes of Centaurus A
  based on the first 10 months of LAT observations (Figure from
  \citet{Abdo2010h}).}
\label{fig:cenaspectrum}
\end{figure*}

The VERITAS observations were prompted by flaring seen in 2009 (Figure
\ref{fig:ngc1275lightcurve}) along with hardening in GeV
\citep{Acciari2009a}.  There was another large flare seen in 2010 along
with a VHE detection ($\sim$60 - 400 GeV) by MAGIC also prompted by
the detection of GeV flaring by the LAT \citep{DonatoD.2010}.  It is
thought that these flares might correlate with flares in the radio and
it will be important in the future to trigger TeV observations off of
radio/GeV flares since no VHE emission has been seen during low
states.

\subsection{Centaurus A}

Centaurus A at a distance of 3.7 Mpc is the nearest radio galaxy and
contains some impressive (10 degrees in extent) radio lobes.  The LAT
has detected emission from both the lobes and the core
\citep{Abdo2010c,Abdo2010h} which indicated that inverse Compton
emission was the source of the MeV/GeV emission in the lobes (Figure
\ref{fig:cenaspectrum}).  The basic understanding is that the
$\gamma$-ray emission is scattered CMB and EBL photons and this
determination allows the measurement of the lobe magnetic field
($\sim$ 1 $\mu$G, near equipartition).  The dominant inverse Compton
(determined from the ratio of the energy desity of the CMB and the
lobe magnetic field: U$_{\rm cmb}$/U$_{\rm b}$ $\sim$ 10) component
indicates that the magnetic field is lower than that seen in other
radio sources.  Looking at the model in Figure \ref{fig:cenaspectrum}
it is apparent that there should be detectable hard X-ray emission but
no evidence of such has been found \citep{Beckmann2011}.  Archival hard
X-ray observations from SAS-3 \citep{Marshall1980} are shown as upper
limits in Figure \ref{fig:cenaspectrum}.

In addition to modelling the emission mechanism of the $\gamma$-ray
photons from the lobes of Cen A, the initial observations allowed the
LAT team to probe the EBL since the inverse Compton/EBL interaction
dominates above 1 GeV.  Unfortunately, the statistics required to
differentiate models is greater than that allowed by the inital 10
month data set \citep{Abdo2010h}.  Deeper analysis using the full 3
year data set is underway.

One of the most exciting developments in VHE astrophysics over the
last several years is the opening up of new populations with long-term
observations and novel analysis techniques.  Centaurus A is one of
those sources.  It was detected in a long (120 hour) observation by
HESS \citep{Aharonian2009}.  The interesting thing is that the VHE and
HE spectra are barely consistent with each other (see Figure
\ref{fig:cenavhe}).  The understanding of this issue is unknown at the
moment but one thing to note is that if FR I's are the parent
population of blazars, than an SSC model should fit the SED but a
simple SSC cannot explain the VHE emission (see Figure
\ref{fig:cenavhe}).  If one allows for the optical and radio emission
to come from a different component than the high energy emission, an
SSC scenario can explain the X-ray to VHE emission \citep{Lenain2008}.

\subsection{M87}

\begin{figure}
\includegraphics[width=65mm]{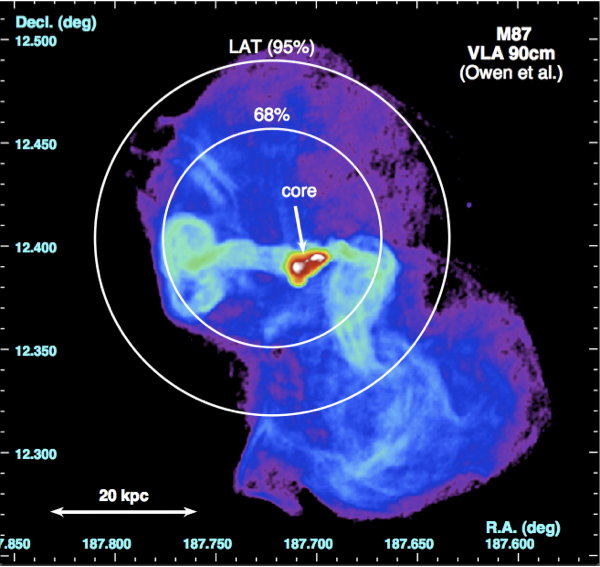}
\caption{VLA image of M87 showing the core and the large radio halo.
  The overlaid circles indicate the LAT 68\% and 95\% containment.
  Many structures are resolved in the radio that cannot be
  disentangled in $\gamma$-rays (Figure from \citet{Abdo2009b}).}
\label{fig:m87skymap}
\end{figure}

The view of M87 in the radio is impressive.  At 90 cm, the jet
outflows terminate in a halo roughly 80 kpc from the core.  At the
scale of a few kpc you can see several knots in X-rays, optical and
radio wavelengths.  M87 is the only non-blazar AGN detected by the
previous generation of VHE instruments (a strong hint of emission at
the 4 sigma level was seen by HEGRA \citep{Aharonian2003b}). Since it
is close (16 Mpc ) you do not have to worry about EBL attenuation and
you can resolve the impressive jet structure in the radio.  The mass
of the central black hole is assumed to be $(3 - 6) \times 10^9 {\rm
  M}_\odot$.  Due to the PSF of VHE and HE instruments, the jet
structure of M87 cannot be resolved at the highest energies (see
Figure \ref{fig:m87skymap}) but by leveraging multiwavelength
observations, flares can be associated with specific knots and other
emission regions.  The LAT emission can be adequately modeled using a
1 zone SSC model assuming a moderate jet beaming of 2 - 4
\citep{Abdo2009b}.

\begin{figure}
\includegraphics[width=65mm]{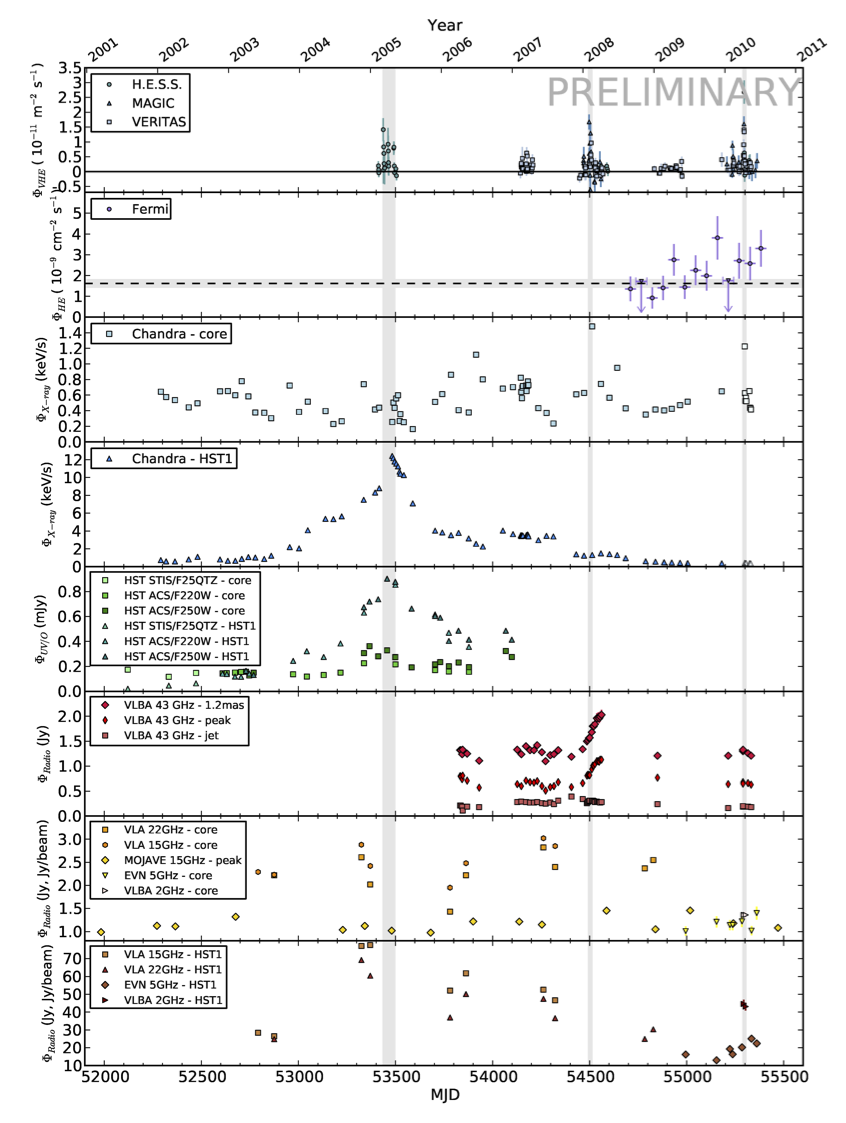}
\caption{Full 10 year multiwavelength lightcurve of M87 (Figure from
  \citet{Abramowski2011}).}
\label{fig:m87lightcurvelong}
\end{figure}

The history of determining the source of the VHE emission is muddled.
Figure \ref{fig:m87lightcurvelong} shows the full 10 year lightcurve
of M87 plotting joint observation campaigns (synchronized and ToO) of
VERITAS, MAGIC, HESS, Fermi, Chandra, HST, VLA and VLBA
\citep{Abramowski2011}.  The coordination and effort needed to produce
such a figure is impressive.  The first flare at VHE (indicated by the
first vertical grey line) was seen in 2005 and it coincided with
flaring at other wavelengths in the knot HST-1, located more than 120
pc from the core.  The conclusion of these observations was the VHE
emission was most likely originating in that knot
\citep{Aharonian2006c}.  It is important to note that when either the
HST-1 knot or the core is X-ray bright, it contaminates the flux of
the other feature.

\begin{figure}
\includegraphics[width=65mm]{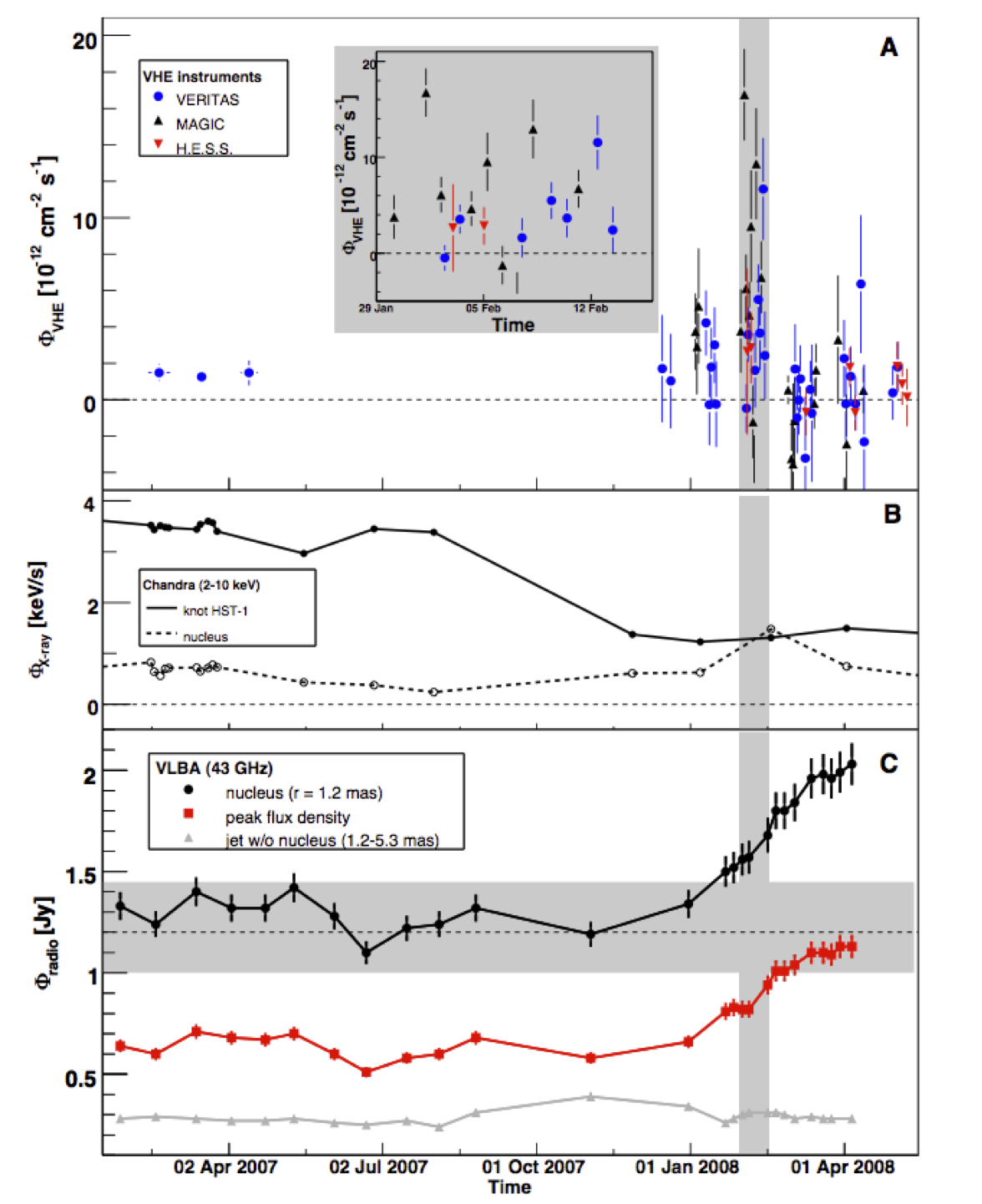}
\caption{Multiwavelength lightcurve of M87 during the 2008 flaring
  period (Figure from \citet{Acciari2009b}.).}
\label{fig:m87lightcurve}
\end{figure}

In 2008 a large flare was seen by all three major VHE instruments
(VERITAS, MAGIC and HESS) and a dedicated multiwavelength campaign was
initiated (Figure \ref{fig:m87lightcurve}).  During this flare, the
VHE emission rose to 10\% of the Crab Nebula's flux, the largest flux
from this galaxy to date and emission was seen to rise in the X-ray
and radio for the core.  The conclusion of these observations was that
the emission originated from the core region of M87 and not from any
knots further down the jet \citep{Acciari2009b}.

The story is further complicated by a large (20\% Crab) flare seen by
VERITAS in 2010 \citep{Ong2010}.  This incident was also observed with the
LAT, VLBA and Chandra.  Even though the core is seen to be brightening
in the X-ray, no increase in flux is seen in the GeV range and the
radio does not exhibit the same behavior as in previous flares.
However, the brightness of the flare in the VHE allows for detailed
studies of the spectral state of M87 during the flaring period.

\section{Conclusions}
It should be apparent that there is still a lot to learn before we can
understand the VHE and HE emission from misaligned blazars (and by
extension AGN in general).  It should also be clear that the only
reason we know as much as we do at this point is due to the amazing
amount of effort put into multiwavelength studies.  Without a full
view of the SED, we could not adequately model these systems. Without
a full view of the lightcurve we could not try and understand the
location of the emission.  It is vitally important to continue
multiwavelength studies from the optical all the way to VHE.

\bigskip 
\begin{acknowledgments}
VERITAS is supported by grants from the U.S. Department of Energy
Office of Science, the U.S. National Science Foundation and the
Smithsonian Institution, by NSERC in Canada, by Science Foundation
Ireland (SFI 10/RFP/AST2748) and by STFC in the U.K. We acknowledge
the excellent work of the technical support staff at the Fred Lawrence
Whipple Observatory and at the collaborating institutions in the
construction and operation of the instrument.

The \textit{Fermi} LAT Collaboration acknowledges support from a
number of agencies and institutes for both development and the
operation of the LAT as well as scientific data analysis. These
include NASA and DOE in the United States, CEA/Irfu and IN2P3/CNRS in
France, ASI and INFN in Italy, MEXT, KEK, and JAXA in Japan, and the
K.~A.~Wallenberg Foundation, the Swedish Research Council and the
National Space Board in Sweden. Additional support from INAF in Italy
and CNES in France for science analysis during the operations phase is
also gratefully acknowledged.
\end{acknowledgments}

\bigskip 

\end{document}